\documentclass[10pt, conference]{IEEEtran}
\usepackage{subfigure, epsfig, color}
\usepackage{amsmath}
\usepackage{amsthm}

\begin{document}

\title{ALOHA-NOMA for Massive Machine-to-Machine IoT Communication}
\author{
  \IEEEauthorblockN{
    Eren Balevi, \textit{Member, IEEE}, 
		Faeik T. Al Rabee, \textit{Student Member, IEEE}, and
    Richard D. Gitlin, \textit{Life Fellow, IEEE}\\}
    \IEEEauthorblockA{Department of Electrical Engineering, University of South Florida \\
		Tampa, Florida 33620, USA\\
      erenbalevi@mail.usf.edu, faeiktayseer@mail.usf.edu, richgitlin@usf.edu}
}
\maketitle 
\normalsize
\begin{abstract}
This paper proposes a new medium access control (MAC) protocol for Internet of Things (IoT) applications incorporating pure ALOHA with power domain non-orthogonal multiple access (NOMA) in which the number of transmitters are not known as $\textit{a priori}$ information and estimated with multi-hypothesis testing. The proposed protocol referred to as ALOHA-NOMA is not only scalable, energy efficient and matched to the low complexity requirements of IoT devices, but it also significantly increases the throughput. Specifically, throughput is increased to $1.27$ with ALOHA-NOMA when $5$ users can be separated via a SIC (Successive Interference Cancellation) receiver in comparison to the classical result of $0.18$ in pure ALOHA. The results further show that there is a greater than linear increase  in throughput as the number of active IoT devices increases.  
\end{abstract}

\begin{IEEEkeywords}
ALOHA, NOMA, M2M Communication, IoT.
\end{IEEEkeywords}

\section{Introduction}
The rapid growth of both the number of connected devices and the data volume that is expected to be associated with the Internet of Things (IoT) applications has increased the popularity of Machine-to-Machine (M2M) type communication within 5G wireless communication systems \cite{Tehrani}. M2M communication without human intervention, which constitutes a significant portion of the IoT, leads to a complete rethinking of the medium access control (MAC) layer. In M2M communication, tens of thousands of low complexity IoT devices can transmit to a gateway. Accordingly, a novel MAC protocol that is scalable, energy efficient and has high throughput is highly desirable. This MAC protocol must be compatible with the low complexity requirements of IoT devices, which have limited battery and memory, as well.

Contention-free MAC protocols such as TDMA, FDMA or CDMA cannot provide high throughput to meet the demands of large number of IoT devices due to control overhead and unused empty slots, and furthermore they are not scalable, though they can prevent collisions. On the other hand, contention based protocols such as CSMA/CA used in IEEE 802.11 perform well only for small networks and they do not have sufficient throughput for large scale networks due to collisions. The same issues are valid for the familiar ALOHA and slotted ALOHA protocols. Moreover, CSMA/CA is energy inefficient, since it requires continuous channel monitoring and there is a significant overhead due to control packets that are not compatible with the limited battery requirements of IoT devices. Furthermore, slotted ALOHA protocols or their variants such as diversity slotted ALOHA \cite{Choudhury} require tight synchronization of the IoT devices in time domain, which can belong to different providers within the framework of heterogeneous IoT networking.

The main aim of this paper is to propose a scalable, energy efficient and high throughput MAC protocol for M2M communication, which is needed as clearly stated in \cite{Rajandekar}. Bearing in mind the simplicity of ALOHA, and the superior throughput of non-orthogonal multiple access (NOMA) \cite{Saito} and its ability to resolve collisions via use of successive interference cancellation (SIC) receiver makes ALOHA-NOMA a good candidate for a MAC protocol that can be utilized for low complexity IoT devices. It is worth noting that the main impediments of ALOHA, which are the low throughput and high collision rate, can be overcome with NOMA. The proposed ALOHA-NOMA protocol is a promising method for not requiring any scheduling in which all IoT devices transmit to the gateway at the same time on the same frequency band but also it is energy efficient such that the devices are not obliged to listen to the channel, and has high throughput as will be demonstrated in this paper.

There are many studies that address MAC layer issues in M2M communication, e.g., see \cite{Rajandekar} and references therein. Among those, the combination of slotted ALOHA with an interference cancellation receiver proposed in \cite{Casini}-\cite{Gallego} are the closest protocols to the proposed ALOHA-NOMA in this paper. However, there are salient differences in the proposed ALOHA-NOMA protocol with the prior art. First, ALOHA-NOMA includes pure ALOHA instead of slotted ALOHA, because slotted ALOHA requires synchronization of hundreds or thousands of IoT devices in time domain. Second, the power levels of IoT devices in ALOHA-NOMA are quite important \cite{Gitlin} to resolve the collisions in the gateway by the SIC receiver \cite{Wang} and they are adjusted by cooperating with the gateway.

The contributions of this paper are $3$-fold. First, a novel scalable, energy efficient, high throughput MAC protocol is proposed to be utilized for IoT applications that have low complexity devices. Following that, a dynamic frame structure that provides great flexibility to accommodate the changing number of IoT devices is defined compatible with the proposed protocol. Finally, the superiority of the proposed ALOHA-NOMA is proven in terms of throughput. 

The paper is organized as follows. Section \ref{Protocol} discusses the ALOHA-NOMA protocol, and Section \ref{dynamic} proposes a frame structure for ALOHA-NOMA. The superiority of the proposed method regarding throughput with respect to the pure ALOHA is proven in Section \ref{Throughput} with its numerical results given in Section \ref{numerical}. The paper ends with the concluding remarks in Section \ref{conclusions}.

\section{ALOHA-NOMA Protocol For IoT Applications} \label{Protocol}
It would seem interesting that MAC layer protocol has not been standardized for IoT applications, while going from theory to practice. The main reason for this can be the contradiction of the low complexity requirements of IoT devices with the high throughput needs. The synergistic combination of the low complexity ALOHA protocol with the promising high throughput feature of NOMA can be an appealing MAC protocol for IoT applications dubbed as ALOHA-NOMA. Note that the growing number of demands makes the state-of-art orthogonal multiple access (OMA) methods insufficient regarding network throughput in which NOMA has emerged as a promising solution in 5G networks \cite{Saito}.

The ALOHA protocol, which was proposed nearly $50$ years ago, has appealing features for IoT applications owing to its simplicity in implementation and compatibility with distributed systems. The main bottleneck of ALOHA systems is the low throughput caused by the high number of collisions, which can be addressed by NOMA. Furthermore, the major impediments of M2M communication namely signaling overhead can be minimized by the combination of ALOHA and NOMA. More precisely, the amount of signaling overhead is reduced in the estimation phase of the proposed protocol in which the number of active devices are estimated by the gateway, which is clarified in the next section, by the unique feature of ALOHA and NOMA where each user transmits whenever it wants. Accordingly, the overhead to establish a connection between the IoT device and the gateway can be minimized before communication begins. This issue is further elaborated in Section \ref{dynamic}.

\begin{figure} [!h] 
\centering 
\includegraphics [width=3.5in]{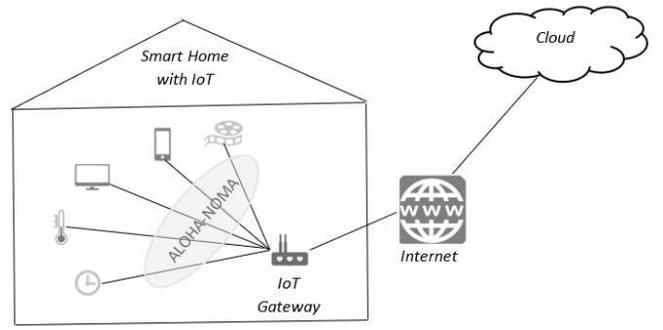}
\caption{A use case for ALOHA-NOMA in the smart home with IoT}\label{fig:model}
\end{figure}

Many IoT devices that are transmitting simultaneously on the same frequency with different power levels to the IoT gateway can be separated via use of SIC receiver employed at the IoT gateway. A sample illustration of this scenario is depicted in Fig. \ref{fig:model} under the notion of smart home with IoT. In this model, IoT devices send their data to the IoT gateway whenever they want using ALOHA-NOMA protocol and the IoT gateway distinguish the signals with SIC receiver. Subsequently, a user can reach these data through internet by using a PC or a mobile in which the data are stored at the registered cloud servers. One of the biggest advantage of ALOHA-NOMA in this topology is to support different heterogenous devices that belongs to different providers without making any network configuration.

The proposed ALOHA-NOMA protocol is scalable owing to the unique properties of ALOHA and NOMA where any node can transmit whenever it wants so that any number of nodes can join or leave the network without any network management involvement before communication. It is an energy efficient protocol due to the fact that a SIC receiver resolves collisions, and thus minimizes retransmission. Lastly, the proposed scheme increases the conventional ALOHA throughput significantly. That is, the normalized throughput increases more than linearly with the total number of users as discussed in Section \ref{Throughput}. The main drawback of the proposed protocol is the increased computational complexity of SIC receiver due to the high number of nodes. However, this can be easily managed, because gateways are much more powerful devices than the low power IoT devices.

\section{Dynamic Frame Structure For ALOHA-NOMA Protocol} \label{dynamic}
One of the main practical challenges in the proposed ALOHA-NOMA protocol to be used in IoT applications is the determination of the proper power levels of IoT devices; otherwise the signals cannot be successfully distinguished at the gateway by a SIC receiver. Indeed, the adjustments of power levels are the only control that must be done for IoT applications before ALOHA-NOMA information transfer begins. To address this challenge, a dynamic frame structure with great flexibility in the changing number of devices is designed, i.e., when a group of IoT devices joins or leaves the network the same frame structure is employed. Such a scheme provides great flexibility in adapting to changing network environments. This structure is opposite to that of TDMA or FDMA in which a new user arrival can completely change the overall frame structure such the additional user must be assigned at least one slot within the frame.

\begin{figure} [!h] 
\centering 
\includegraphics [width=3.5in]{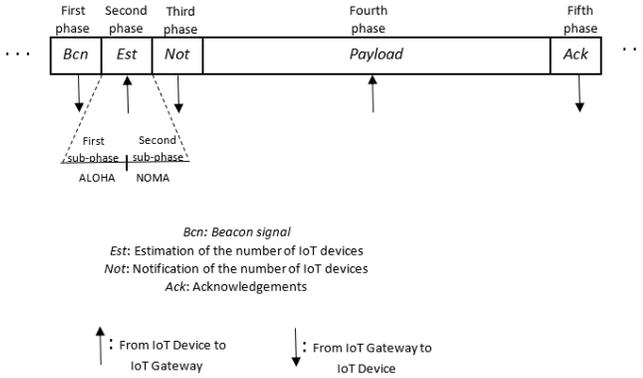}
\caption{The proposed frame structure}\label{fig:frame}
\end{figure}

The proposed frame structure is periodic and basically composed of $5$ phases. An illustration of the proposed periodic frame structure is given in Fig. \ref{fig:frame}. Accordingly, the gateway first transmits a beacon signal. Next, the IoT devices with packets to transmit send ``dummy" (without data content) packets to help the gateway estimate the total number of active devices in the medium. It follows that the number of IoT devices are estimated at each period via multi-hypothesis testing \cite{Shaffer}, and the SIC receiver is implemented\footnote{We denote a SIC receiver that can process $N$ signals as SIC($N$) and we refer to $N$ as the SIC degree and $N$ is adjusted to agree with the result of the multi-hypothesis test described above.} to decode this number of different packets. Note that detecting the number of signals using multiple hypothesis testing was previously shown in \cite{Chung}. Third, the total number of devices is broadcast to the transmitters, and each IoT device adjust its transmission power properly. Notice that each device has a number/identity that maps to the appropriate power level. Fourth, all IoT device transmit their packets to the gateway, and last the detected messages are acknowledged. The fifth phase, an ACK packet, can contain the unique IoT device numbers corresponding to successfully decoded packets so that each device can understand whether its packet is successfully received. Note that this frame structure is preserved when the number of IoT devices changes, which provides great flexibility, because the length of phases is independent from the number of IoT devices. Although one can consider that the first three phases and the ACK decrease throughput efficiency, they are considerably shorter than the fourth phase or payload. 

To give more detail on the protocol, the gateway initially transmits a beacon signal and in the second phase, all IoT devices that send packets in this frame transmit simultaneously on the same frequency to the IoT gateway using an ALOHA  protocol. Then, the IoT gateway estimates the number of IoT devices by multi-hypothesis testing and the SIC receiver is set to decode this number of IoT devices using NOMA. Notice that it would lead to easier implementation if all IoT devices are registered to the gateway instead of multi-hypothesis testing, however, this may significantly increase the length of control phase and thus decrease the payload or throughput considering the large number of IoT devices. A popular method in multi-hypothesis testing that can be used to find the number of IoT devices at the gateway is based on the Bonferroni Inequality \cite{Shaffer}. More precisely, $M$ independent null hypotheses are tested such as $H_1$, $H_2$, $\cdots$, $H_M$ where $H_i$ is the event that $i^{th}$ user exists. Each of the hypothesis has corresponding $p$-values as $p_1, p_2, \cdots, p_M$, respectively. These $p_i$ values are Gaussian random variables with mean $E_i$ and variance $\sigma^2$ where $E_i$ is the average signal power. The number $N$ is the total number of true null hypothesis or the total number of IoT devices in the medium such that $N\leq M$. The value of $N$ is unknown at the gateway and estimated probabilistically. Accordingly, the probability of having $N$ IoT devices can be specified as
\begin{equation}
P(N\ number\ of\ IoT\ devices)=(1-\alpha)^N\alpha^{M-N}
\end{equation}
where $\alpha$ is determined using the Bonferroni Inequality given by 
\begin{equation}
P\left(\bigcup_{n=1}^M\left(p_i\leq \frac{\alpha}{M}\right)\right)\leq \alpha.
\end{equation}

Once the total number of IoT devices, $N$, is estimated, the SIC receiver at the gateway decodes $N$ strongest signal. Assuming IoT devices send their address/ID in the “dummy” packet, the gateway decodes these packets and broadcasts the number of devices with these addresses/IDs in the third phase. The devices who do not get detect their address/ID, do not transmit a payload. Those devices that detect their address/ID from the gateway message change their power according to the power back-off scheme as proposed in \cite{Zhang}. In particular, they change their power level with $n\Delta$ where $n$ is a uniform random variable between -$N$ and $N$, and $\Delta$ is a pre-determined value. That is, each of the IoT devices randomly selects a number $n$ and changes its transmission power to enhance the quality of the SIC receiver where the transmitters must have different power levels. The active IoT devices that do not detect their address/ID from the gateway can increase their transmit powers slightly in the next round. This will affect the outcome of the multi-hypothesis testing as well. In the fourth phase, the payload information is sent to the gateway by all the IoT devices that detected their address/ID from the gateway. Notice that the fourth phase of the proposed frame structure can be considered as pure NOMA. The successfully detected packets are acknowledged in the last phase. This procedure repeats periodically.  

\section{Throughput Of ALOHA-NOMA} \label{Throughput}
One of the most critical parts of the proposed frame structure depends on the determination of the number of IoT devices in the second-phase. If only pure ALOHA was used in the second phase, the performance would be unsatisfactory since only 18\% of the IoT devices can be detected at a time on average. To address this problem, the ALOHA protocol is replaced with ALOHA-NOMA in the second phase of the proposed frame structure. Accordingly, the number of devices are first estimated according to the superposed signal strength based on multi-hypothesis testing as was done in \cite{Chung}. Note that in this phase the aim is not to detect the packets, but only to estimate the number of devices. Once the number of active devices is estimated, the SIC receiver is set to decode this number of signals and the packets are detected. It is worth emphasizing that at each round the power level of each device randomly and independently changes, and thus SIC receiver can successfully work relying on these power differences. These two consecutive parts inside the second phase are denoted as ALOHA and NOMA, respectively, which explains the reason of referring to this protocol as ALOHA-NOMA

The main aim of this section is to specify the throughput increase of ALOHA-NOMA with respect to pure ALOHA. The ALOHA throughput can be considerably increased using NOMA, which will allow simultaneous users to successfully communicate with the gateway during the vulnerable period. For a SIC receiver that can successfully detect $N$ transmissions, that is a SIC($N$), successful reception will occur when there are $N$ or fewer arrivals (transmissions) that occurs with probability
\begin{equation}
P(successful\ transmission)=P(N\ or\ fewer\ arrivals).
\end{equation}

Assuming a large number of IoT devices, the Poisson distribution is a reasonable communication model and the probability of $i$ transmissions during the vulnerable time period using NOMA with $N$ power levels is given by 
\begin{equation}\label{4}
P(i\ transmissions)=\frac{(2gT)^ie^{-2gT}}{i!}
\end{equation}
where the parameter $g$ represents the arrival rate (packets/ second) and $i = 0,1,2, \cdots$ Since the probability of ''success” is the probability of $N$ or fewer arrivals, the normalized throughput, which can be defined as the time fraction during which the useful information can be carried on the channel, is given as
\begin{equation}\label{5}
S_{th}=\sum_{i=1}^N\frac{(2G)^ie^{-2G}}{2(i-1)!}=\frac{e^{-2G}}{2}\sum_{i=1}^N\frac{2^iG^i}{(i-1)!}
\end{equation}
where $G = gT$ is the normalized offer load. 

In order to find the relation between the maximum throughput of ALOHA-NOMA for different values of $N$, (\ref{5}) is used to compute the maximum throughput for each  value of $N$. For $N=1$, the maximum throughput is defined by differentiating (\ref{5}) with respect to $G$ and equating to zero as
\begin{equation}\label{6}
\frac{\partial S_{th}}{\partial G}=\frac{(-4e^{-2G}G)+(2e^{-2G})}{2}=0. 
\end{equation}
Solving for $G$ in (\ref{6}) gives the familiar result $G = 1/2$. Substituting $G=1/2$ in (\ref{5}) for $N = 1$, gives the familiar result of $0.18$.

The same process may be applied with $N = 2$. The throughput, using (\ref{5}) with $N = 2$, is 
\begin{equation}\label{7}
S_{th}=\sum_{i=1}^2\frac{(2G)^ie^{-2G}}{2(i-1)!}.
\end{equation}
The maximum throughput is obtained by differentiating (\ref{7}) with respect to $G$ and equating the result to zero as
\begin{equation}\label{8}
\frac{\partial S_{th}}{\partial G}=\frac{e^{-2G}}{2}(2+4G-8G^2)=0.
\end{equation}
Solving (\ref{8}), the roots of $G$ are $0.809$ and $-0.309$. Therefore, to find the maximum throughput of ALOHA using NOMA with $N =2$, $G = 0.809$ is substituted in (\ref{7}) and the result is given as
\begin{equation}\label{9}
S_{th\ max\ at\ G=0.8090}=0.42.
\end{equation}
As shown in (\ref{9}), for $N=2$, the maximum (normalized) throughput of ALOHA-NOMA is increased from 18\% to 42\%. For $N = 3$ the maximum throughput is 
\begin{equation}\label{10}
S_{th\ max\ at\ G=1.1348}=0.6856.
\end{equation}
This procedure can be applied for any number of $N$ to determine the maximum throughput for the ALOHA-NOMA system.

\section{Numerical Results} \label{numerical}
\begin{figure} [!h] 
\centering 
\includegraphics [width=3.5in]{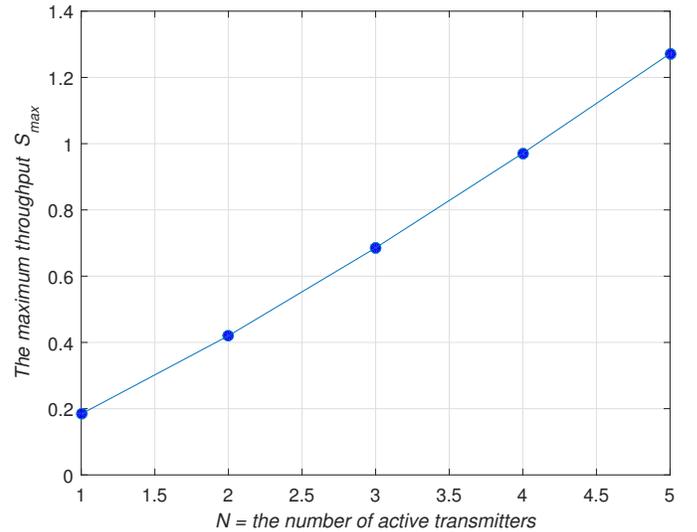}
\caption{ALOHA-NOMA maximum throughput as a function of N}\label{fig:ICC1}
\end{figure}

\begin{figure} [!h] 
\centering 
\includegraphics [width=3.5in]{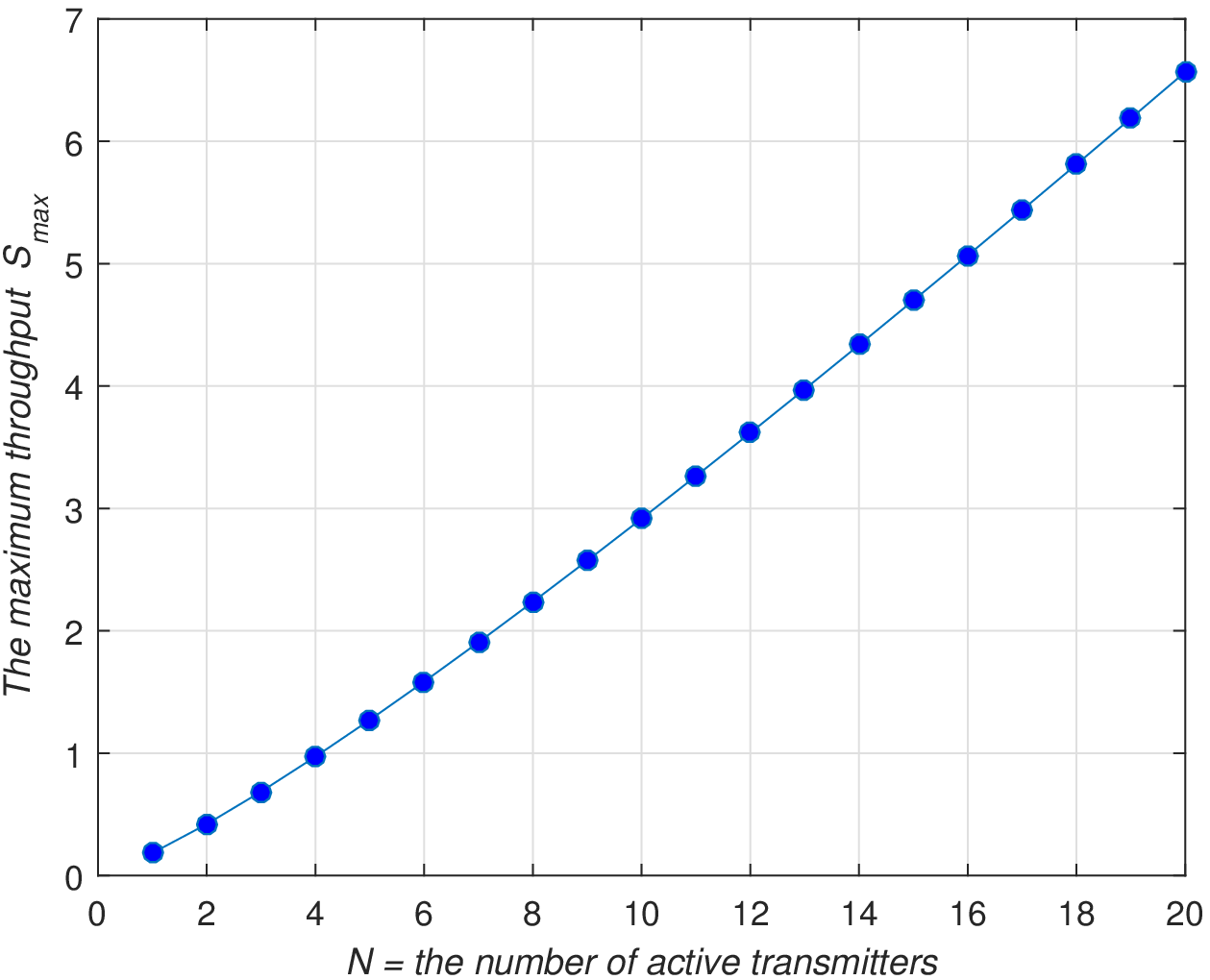}
\caption{ALOHA-NOMA maximum throughput as a function of N}\label{fig:ICC2}
\end{figure}

\begin{figure} [!h] 
\centering 
\includegraphics [width=3.5in]{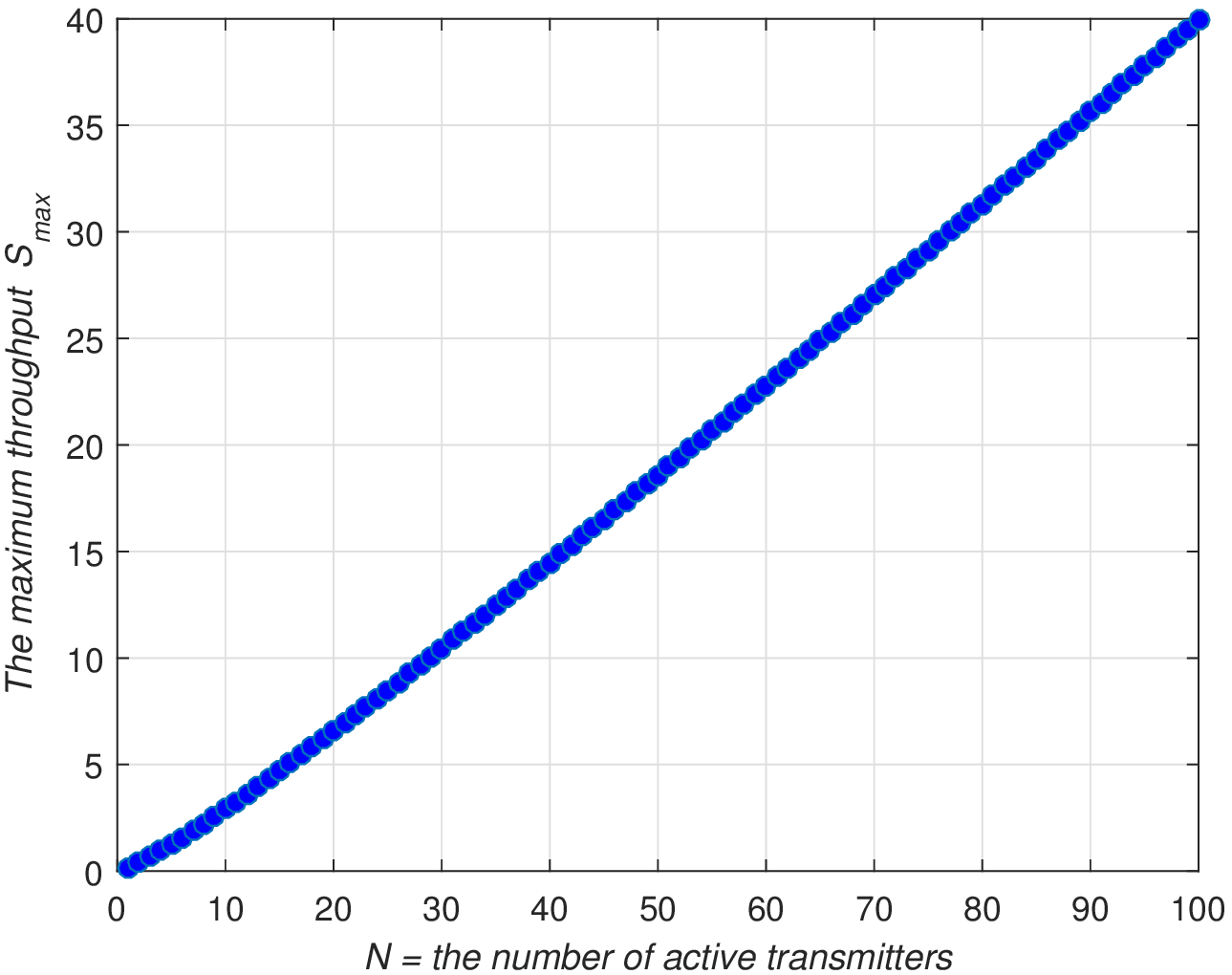}
\caption{ALOHA-NOMA maximum throughput as a function of N}\label{fig:ICC3}
\end{figure}

The throughput analysis of ALOHA-NOMA is numerically evaluated to give more insight for different values of $N$. The performance measure is the maximum throughput that can be found using (\ref{5}). That is, (\ref{5}) is evaluated for different values of $N$. As shown in Fig. \ref{fig:ICC1}, the normalized throughput of ALOHA-NOMA increases with a slope that is greater than linear with $N$ ranging from $1$ to $5$. Notice that the familiar result of ALOHA throughput, which is $0.18$, is observed at $N = 1$ in Fig. \ref{fig:ICC1} for the ALOHA-NOMA system. The normalized throughput is investigated for a large number of active transmitters, for $N = 20$ and $N = 100$ in Fig. \ref{fig:ICC2} and Fig. \ref{fig:ICC3}, respectively. Note that Fig. \ref{fig:ICC2} and Fig. \ref{fig:ICC3} are the generalization of Fig. \ref{fig:ICC1} for high values of $N$. These results depict that the maximum throughput increases with a greater than linear slope when the number of active IoT users, $N$, increases.

\section{Conclusions} \label{conclusions}
This paper is directed towards a novel MAC layer protocol for IoT applications. ALOHA-NOMA, a simple, easy to implement, distributed, and thus scalable protocol is proposed that is compatible with the low complexity requirements of IoT devices. Although ALOHA-NOMA is energy efficient, since the SIC receiver at the gateway will minimize the retransmissions of IoT devices and does not need to listen the channel continuously, its throughput efficiency can be degraded by the fact that it may be difficult to distinguish multiple signals that transmit at the same time on the same frequency band, which is necessary to determine the number of active IoT devices. To address this challenge, a dynamic frame structure is designed that is robust to the changing number of devices,  so that estimation of the number of active IoT devices transmitted in a ALOHA is accomplished  using multi-hypothesis testing that detects the packets of IoT devices using a NOMA SIC-based receiver. This estimation phase is much smaller in duration than the payload, and hence it does not decrease the efficiency of the throughput. With correct operation of the multi-hypothesis testing procedure and successful SIC detection, ALOHA-NOMA significantly improves the throughput performance with respect to the pure ALOHA, e.g., a SIC receiver that separates $5$ signals can boost the throughput of classical ALOHA from $0.18$ to $1.27$ and with $100$ active IoT devices the throughput is increased (at a greater than linear rate) to $40$.

\section*{Acknowledgement}
Faeik Al Rabee is supported by Al-Balqa' Applied University (BAU), Jordan. The statements made herein are solely the responsibility of the author.


\begin{thebibliography}{1} 

\bibitem{Tehrani}
M. N. Tehrani, M. Uysal, and H. Yanikomeroglu, 
``Device-to-device communication in 5G cellular networks: Challenges, solutions, and future directions",
\emph{IEEE Commun. Mag.}, vol. 52, no. 5, pp. 86–92, May 2014.

\bibitem{Choudhury}
G. L. Choudhury and S. S. Rappaport, 
``Diversity ALOHA - A random access scheme for satellite communications",
\emph{IEEE Trans. Commun.}, vol. 31, pp. 450–457, Mar. 1983.

\bibitem{Rajandekar}
A. Rajandekar and B. Sikdar, 
``A survey of MAC layer issues and protocols for machine-to-machine communications",
\emph{IEEE Internet of Things Journal}, vol. 2, no. 2, pp. 175-186, Apr. 2015.

\bibitem{Saito}
Y. Saito, Y. Kishiyama, A. Benjebbour, T. Nakamura, A. Li, and K. Higuchi,
``Non-orthogonal multiple access (NOMA) for cellular future radio access",
\emph{in Proc. IEEE Veh. Technol. Conf. (VTC Spring)},  pp. 1–5, Jun. 2013.

\bibitem{Casini}
E. Casini, R. De Gaudenzi, and O. Herrero,
``Contention resolution diversity slotted aloha (CRDSA): An enhanced random access scheme for satellite access packet networks",
\emph{IEEE Transactions on Wireless Communications}, vol. 6, no. 4, pp. 1408–1419, April 2007.

\bibitem{Liva}
G. Liva,
``Graph-based analysis and optimization of contention resolution diversity slotted ALOHA",
\emph{IEEE Transactions on Communications}, vol. 59, no. 2, pp. 477–487, February 2011.

\bibitem{Stefanovic}
C. Stefanovic, K. Trilingsgaard, N. Pratas, and P. Popovski,
``Joint estimation and contention-resolution protocol for wireless random access",
\emph{in IEEE International Conference on Communications (ICC)}, pp. 3382–3387, June 2013.

\bibitem{Ricciato}
F. Ricciato and P. Castiglione,
``Pseudo-random ALOHA for enhanced collision-recovery in RFID",
\emph{IEEE Communications Letters}, vol. 17, no. 3, pp. 608–611, March 2013.

\bibitem{Gallego}
F. V. Gallego, M. Rietti, J. Bas, J. Alonso-Zarate and L. Alonso,
``Performance evaluation of frame slotted-ALOHA with successive interference cancellation in machine-to-machine networks",
\emph{in Proc. 20th Eur. Wireless Conf.}, pp.1–6, May 2014.

\bibitem{Gitlin}
F. Al Rabee, K. Davaslioglu, and R. D. Gitlin,
``The optimum received power level of uplink non-orthogonal multiple access (NOMA) signals",
\emph{IEEE Wireless and Microwave Technology Conference (WAMICON)}, pp. 1-4, April 2017.

\bibitem{Wang}
D. N. C. Tse and P. Viswanath, 
Fundamentals of Wireless Communications,
U.K., Cambridge:Cambridge Univ. Press, 2005.

\bibitem{Shaffer}
J. P. Shaffer,
``Multiple hypothesis testing",
\emph{Annual Review of Psychology}, 46, 561–584, 1995.

\bibitem{Chung}
P.-J. Chung, J. F. Bhme, A. O. Hero, C. F. Mecklenbruker,
``Signal detection using a multiple hypothesis test",
\emph{Proc. 3rd IEEE Sensor Array and Multichannel Signal Process. Workshop}, pp. 221-224, 2004.

\bibitem{Zhang}
N. Zhang, J. Wang, G. Kang, and Y. Liu,
``Uplink non-orthogonal multiple access for 5G Systems",
\emph{IEEE Commun. Lett.}, vol. 20, no. 3, 458–461, Mar. 2016.

\end{thebibliography}
\end{document}